\documentclass[reprint,
amsmath,amssymb,
aps,
]{revtex4-2}

\usepackage{amsmath}    
\usepackage{docs}%
\usepackage{bm}%
\usepackage[colorlinks=true,linkcolor=blue, citecolor=magenta]{hyperref}%
\usepackage[usenames]{xcolor}
\usepackage{xcolor}
\usepackage{graphicx}
\usepackage{pstool}
\usepackage[export]{adjustbox}

\def\av#1{\langle{#1}\rangle}
\def\pti#1{ }
\newcommand{\qq}{\,,\qquad}
\def\be#1{\begin{equation}\label{#1}}
\def\ee{\end{equation}}
\def\ba#1#2{\be{#1}\begin{array}{#2}}
	\def\ea{\end{array}\ee}
\def\={\stackrel{\mbox{\scriptsize def}}{=}}
\def\bra#1{\left({#1}\right)}
\begin{document}
	\title{Transition to thermal equilibrium in a deformed crystal}
	
	\author{A.M. Krivtsov}
	\email{akrivtsov@bk.ru}
	\affiliation{IPME RAS}
	\affiliation{Peter the Great St.~Petersburg Polytechnic University}
	
	\author{A.S. Murachev}
	\email{andrey.murachev@gmail.com}
	\affiliation{Peter the Great St.~Petersburg Polytechnic University}
	\begin{abstract}
	An adiabatic transition between two equilibrium states corresponding to different stiffnesses in an infinite chain of particles is studied. Initially, the chain particles have random displacements and random velocities corresponding to a uniform initial temperature. An instant change of parameters of chain initiates a transient process. Analytical expressions for the chain temperature as a function of time are obtained from the statistical analysis of the dynamics equations. It is shown that the transition process is oscillatory and it converges non-monotonically to a new equilibrium state. Such behavior is usually unexpected for thermal processes. 
	The analytical results are supplemented by numerical simulations.
\end{abstract}
\maketitle

\section{Introduction}
Nonequilibrium thermal processes in solids at nano- and microlevel are currently a subject of intensive research, partly driven by development of nano technologies~\mbox{\cite{Goldstein2007, Golovnev2006, Korobeynikov2012, KrivMoroz2002, Baimova2017,Vikesland18}}.
At microlevel, a transition to an equilibrium state for nearly harmonic crystals is a gradual equalization of the kinetic and potential energies of the atomic motion according to the virial theorem~\cite{Hoower2015, Altenbah, Kuzkin2015, Kuzkin2017,Bass}.
However, this theorem does not describe the processes that occur during the transients.
A macroscopic description of such transients is also challenging since it requires application of special constitutive equations for ultrafast atomic processes.
{Therefore, we develop a simple model that covers analytically both microscopic and macroscopic  descriptions of the transient processes.}

Crystals with simple lattices are convenient models for studying nonequilibrium processes in solids~\cite{Lepri2003, Rieder1967, Kannan2012, Xiong2013}. Numerical simulations~\cite{Allen1987} showed that the process of energies equalization in molecular systems was accompanied by high-frequency oscillations.
In pioneering paper by Klein and Prigogine~\cite{Prig}, the equations of the atomic motion for a one-dimensional harmonic crystal were solved directly and it was shown that the energy oscillations after an instantaneous thermal perturbation were described by the Bessel function of the first kind. In later work~\cite{Krivtsov 2014 DAN}, this problem was solved by analyzing the dynamics equations for the velocity covariances, which allowed to generalize these results for more complex systems, including multidimensional crystals~\cite{Kuzkriv2016, KuzArxiv2017, Murachev2018, GavrKriv2019, Sokolov2019, Berinskii2019}.

In the mentioned works the nanoscale thermal processes were studied separately from the mechanical processes. The advances of the modern technologies bring to attention ultrafast mechanical processes where the speed of mechanical load is comparable or even faster than the speed of the local thermal equilibration in the system.
This condition is fulfilled if the material is rapidly deformed by forces uniformly distributed along the length of the sample.
Such loads occur in nanoscale electronic components of experimental equipment requiring fast magnetic field switches, for example, necessary for condensed matter physics, plasma physics, or inertial confinement synthesis \cite{sen}.
Electric pulses can create distributed electromagnetic loads on circle samples for the time up tens nanoseconds~\cite{Zhang, Morozov}. In \cite{Kruglyak}, the concepts of an ultrafast photodetector capable of converting femtosecond light pulses into electric pulses of the same length is proposed. Paper \cite{sen} demonstrates an all-optical method to generate magnetic field impulses of the order of several tesla over the course of tens of femtoseconds.
In the nearest future it is expected that the experimental electromagnetic impulses can reach a duration less than femtosecond: the femtosecond lasers already exist~\cite{Toyserkani} and attoseconds lasers are underdevelopment \cite{Tzallas,Orfanos,Feng}. Therefore, an analytical study of the impact of the ultrafast mechanical loads on the thermal equilibration is needed to provide a theoretical basis for the upcoming experimental studies.

In the present paper, we study an adiabatic non-equilibrium process analogous to those considered previously~\cite{Kuzkriv2016, KuzArxiv2017, Murachev2018, GavrKriv2019, Sokolov2019, Berinskii2019}. However, this process is initiated by an instantaneous external load instead of instantaneous heating.
The material is deformed by forces uniformly distributed along the length of the sample~\cite{Zhang, Morozov} hence for this case the orders of velocity of mechanical and thermal processes are the same.
The interaction between mechanical and thermal processes requires accounting non-linearity, therefore $\alpha$-FPU one-dimensional crystal~\cite{FPU_original, FPU_review} is considered.
The non-linearity is assumed to be large enough to cause an adiabatic heating of the crystal, but, on the other hand, is sufficiently small to analysis the resulting energy oscillations in the framework of the harmonic approximation.

The paper is organized as follows. In Sect.\ref{sectModel} the mathematical formulation of the problem is presented.
In Sect.\ref{sect4} the equilibrium value of the kinetic temperature after loading is obtained.
In Sect.\ref{SectDynamics} it is shown that the crystal temperature oscillates during the transition process and an analytical expression that describes these oscillations is obtained.  In the particular system considered, the kinetic and potential energies gradually become equal over time (in agreement with the virial theorem). The analytical solution is compared with a numerical one.
In Sect.\ref{sectEx} an estimation of the temperature jump for the real crystal is represented.

\section{Formulation of the problem}\label{sectModel}
A model of an infinite one-dimensional crystal --- chain of point masses connected by unmasses springs is considered.
It is assumed that the chain  particles interact only with their nearest neighbours. 
In thermodynamic equilibrium state particle velocities $v_n$ and bond deformations $\varepsilon_n$ are independent stochastic quantities and all statistical characteristics of the crystal are constant over time. 
To describe the statistical behaviour of the system the following quantities are introduced:
\begin{equation}\label{Energies}
	\begin{split}
		K\=\frac{1}{2}m\av{v^2_n}\qq  
		T\=\frac{2K}{k_B},
	\end{split}
\end{equation}
where  $K$ is the mathematical expectations of kinetic energy,~$T$~is the kinetic temperature, $n=1,2,3,...$ is the index of the particle, $\av{...}$ is the operator of the mathematical expectation,  $v_n$ is the particle velocity, $k_B$~is the Boltzmann constant, and~$m$ is the particle mass.
The statistical characteristics of this system depend on such crystal parameters as particle mass and bond stiffness. Variation of these parameters 
transfers the crystal to a non-equilibrium state and the transition process starts, which eventually brings the crystal to a new equilibrium state. 

One of the natural way to change the bond stiffness of the crystal is a homogeneous crystal deformation by external loading. For example, such loading can be realized by applying distributed electromagnetic forces to a circle samples~\cite{Zhang, Morozov}.  Based on the approach described~\cite{Krivtsov 2014 DAN}, we investigate the evolution of the kinetic temperature of the crystal during this transition process.

In addition to mathematical expectation of kinetic energy $K$, this paper uses mathematical expectation of potential energy $U$:
\begin{equation}\label{pot}
	\begin{split}
		&U(\epsilon+\varepsilon_n)\=\av{\varPi(\epsilon+\varepsilon_n)}
		\qq \varepsilon_n\=u_n-u_{n-1},\\		&\varPi(\varepsilon)\=\frac{C}{2}{\varepsilon^2}
		+\frac{\alpha}{3}{\varepsilon^3},
	\end{split}
\end{equation}
where $\varPi$ is the inter-particle potential, $C$ and $\alpha$ are stiffness of the inter-particle bonds of the first and the second order, $u_n$ is the displacement of the particle from its equilibrium position, and
deformation is a sum of homogeneous deformation $\epsilon$ and stochastic deformation~\mbox{$\varepsilon_n$}.
Homogeneous deformation~$\epsilon(t)$ is applied instantaneously:
\begin{equation}\label{deform}
	\epsilon(t)=
	\begin{cases}
		0, &  t<0\\
		\epsilon, &   t \geq 0,
	\end{cases}
\end{equation}
where  $\epsilon$ is a constant.
The force $F_n$ acting on the particle~\mbox{$n-1$} by the particle $n$ is 
\begin{equation}\label{force}
	F_n\=-F(\epsilon+\varepsilon_n)\qq F(\varepsilon)=-\varPi'(\varepsilon)
	=-C\varepsilon-\alpha\varepsilon^2,
\end{equation}
and equation of particle dynamics  of the crystal is:
\begin{equation}\label{dynamic}
	m{\dot v}_n =F_{n+1}-F_n.
\end{equation}

We consider small non-linearity, therefore the second terms in the formulas for $\varPi(\varepsilon)$~\eqref{pot} and $F(\varepsilon)$~\eqref{force} are small compare to the first one. Since the non-linear term is small, a harmonic (linear) approximation is used to describe the thermal processes. At relatively short times, the harmonic approximation is quite accurate for describing thermal processes in a crystal with small non-linearity~\cite{Kuzkriv2016}.
However, non-zero stiffness of the second order allows taking into account the influence of homogeneous deformation $\epsilon$ on inter-particle potential.

The analytical expression for kinetic temperature at long times after homogeneous deformation can be obtained using the virial theorem and is given in the following section.

	\section{Crystal in the thermodynamic equilibrium}\label{sect4}
	This section contains expressions for the kinetic temperature of the crystal before loading and at large times after loading.	
	
	\subsection{Prior to loading}
	
	At $t<0$ according to~\eqref{deform}, the
	energies are
	\begin{equation}\label{Energies_before_loading}
	\begin{split}
	&K=\frac{1}{2}m\av{v^2_n}\qq U=\frac{C}{2}\av{\varepsilon_n^2}
	+\frac{\alpha}{3}\av{\varepsilon_n^3}
	\end{split}
	\end{equation}
	and the crystal is assumed to be in thermodynamic equilibrium. 	Since only the mathematical expectations of the energies are considered hereinafter, the words ``mathematical expectation'' are omitted.
	Following~\cite{Altenbah} the kinetic energy~\eqref{Energies} of the system can be represented by (see appendix \ref{app0}):
	\begin{equation}\label{virial}
		K=\frac{1}{2}\av{\varepsilon_nF_n}.
	\end{equation}	
	Formula~\eqref{virial} allows one to obtain expressions for the equilibrium kinetic temperature of the crystal before and after the instantaneous deformation, as follows:
	\begin{equation}\label{ver}
	m\av{v^2_n}=\av{\varepsilon_n F_n}
	={C}\av{\varepsilon_n^2}
	+{\alpha}\av{\varepsilon_n^3}.
	\end{equation}
	We neglect the small term ${\alpha}\av{\varepsilon_n^3}$ in equation~\eqref{ver} and obtain
	the kinetic temperature of the crystal prior to loading
	\begin{equation}\label{T0}
	k_BT_0={m}\av{v_n^2}={C}\av{\varepsilon_n^2}.
	\end{equation}
In the next subsection, we consider the thermodynamic quantities after loading.	

	\subsection{After loading}
	At $t\geqslant 0$, the
	kinetic and potential energies of the crystal:
	\begin{equation}\label{Energies_after_loading}
	\begin{split}
	&K=\frac{1}{2}m\av{v^2_n}\qq U=\varPi(\epsilon)+U_T\qq \\
	&U_T=
	\frac{1}{2}\bra{C+2\alpha\epsilon}\av{\varepsilon_n^2}+\frac{1}{3}\alpha \av{\varepsilon_n^3},
	\end{split}
	\end{equation}
    where $U_T$ is the thermal part of the potential energy.
	In the state of thermodynamic equilibrium the kinetic energy $K$ and the thermal part of the potential energy~$U_T$ are equal, therefore expression for kinetic temperature~\eqref{Energies} is: 
	\begin{equation}\label{Tempfr}
	\begin{split}
		\left.k_BT\right\vert_{t\to\infty}=2K=U_T+K.
	\end{split}
\end{equation}
	Note that $U_T+K$ is constant for $t\geq 0$ in the transition process. An expression for this sum can be obtained by substitution to~\eqref{Tempfr} of the initial values of the corresponding quantities:
	\begin{equation}\label{Tempfr1}
	\begin{split}
		k_BT=\left.\frac{1}{2}m\av{v^2}\right\vert_{t=0}+\left.\frac{1}{2}(C+2\alpha\epsilon)\av{\varepsilon_n^2}\right\vert_{t=0},
	\end{split}
	\end{equation}	
	where term~$\alpha \av{\varepsilon_n^3}/3$ is omitted.
	Substituting expressions~\eqref{T0} into~\eqref{Tempfr1} yields:
	\begin{equation}\label{Tfinal}
	\begin{split}
		\left. T\right\vert_{t\to\infty}=T_0\bra{1+\frac{\alpha\epsilon}{C}}.
	\end{split}
	\end{equation}
	
	From expression~\eqref{Tfinal} it follows that the change in kinetic temperature $T$, in a first approximation, is proportional to deformation $\epsilon$. 
		
	\section{Dynamics of the transition process}\label{SectDynamics}
	{In this section we obtain an expression for the temperature as a function of time.}
	Substituting expression~\eqref{force} into~\eqref{dynamic} we obtain the following equation of motion:
	\begin{equation}\label{dc00}
	m\dot v_n =C\Delta \varepsilon_n+2\alpha\epsilon\Delta \varepsilon_n+\alpha\Delta \varepsilon_n^2\qq
	\dot \varepsilon_n=\Delta v_n.
	\end{equation}
	Term $\alpha\Delta \varepsilon_n^2$ can be neglected in the case of small deformations. Thus equations \eqref{dc00} become linear:
	\begin{equation}\label{dynamictgeq0}
	\dot v_n =\omega^2\Delta \varepsilon_n\qq \dot \varepsilon_n=\Delta v_n
	\end{equation}
	where $\omega\=\sqrt{\frac{C+2\alpha\epsilon}{m}}$. Note that sum $C+2\alpha\epsilon$ plays a role the first order stiffens after external loading.
	The initial conditions for the system~\eqref{dynamictgeq0} are determined from
	equation~\eqref{T0}:
	\begin{equation}\label{inicond}
	\left.v_n\right|_{t=0}=\sqrt{\frac{k_BT_0}{m}}\rho_n\qq \left.\varepsilon_n\right|_{t=0}=\sqrt{\frac{k_BT_0}{C}}\varrho_n\qq
	\end{equation}
	where $\rho_n$ and $\varrho_n$ are independent random numbers with zero mathematical expectation and unit variance.
	The initial value problem \eqref{dynamictgeq0}-\eqref{inicond} describes the stochastic dynamics of the chain particles. Then the kinetic temperature of the crystal as a function of time can be found using the covariance analysis approach~\cite{Krivtsov 2014 DAN, GavrKriv2019}, by introducing generalized energies
	\begin{equation}\label{GenEn}
	\begin{split}
	&{\cal K}_k\=\frac{1}{2}m\av{{v}_n{v}_{n+k}}\qq
	 {\cal U}_k\=\frac{1}{2}m\omega^2\av{{\varepsilon}_n{\varepsilon}_{n+k}}\qq\\
	&{\cal L}_k\=  {\cal K}_k-  {\cal U}_k,
	\end{split}
	\end{equation}
where ${\cal K}_k$ and ${\cal U}_k$ are generalized kinetic and potential energies, {and} ${\cal L}_k$ is generalized Lagrangian.
	Differentiating of~\eqref{GenEn} and equations of motion \eqref{dc00} lead to the following initial problem for the generalized Lagrangian:
	\begin{equation}\label{LagProb}
	\begin{split}
	&{\ddot  {   {\cal L}}}_k=4\omega^2(  {   {\cal L}}_{k-1}-2  {   {\cal L}}_k+ {   {\cal L}}_{k+1})\qq\\
	&t=0:\quad  {   {\cal L}}_k=- \frac{T_0k_B\alpha\epsilon}{C}\delta_k\qq \dot  {   {\cal L}}_k=0,
	\end{split}
	\end{equation}
	where  $\delta_k=1$ for $k=0$ and $\delta_k=0$ otherwise.
	Solution of a similar initial problem is obtained in~\cite{Krivtsov 2014 DAN}. According to that solution, generalized Lagrangian oscillates with a monotonically decreasing amplitude:
	\begin{equation}\label{sol cl}
	  {\cal L}_k=- \frac{T_0k_B\epsilon\alpha}{C} J_{2k}(4\omega t),
	\end{equation}
	where $J_k(x)$ is the Bessel function of $k$-th order~\cite{Abramowitz}.
	
	Kinetic energy~$K$ and thermal part of the potential energy $U_T$
	are equal to the generalized kinetic and potential energies with zero indexes, therefore
	kinetic temperature~$T$ can be expressed as follows:
	\begin{equation}\label{Temp}
	k_BT={\cal L}_0+K+U_T.
	\end{equation}
Then the kinetic temperature of the crystal as a function of time can be found using formulas~\eqref{Tfinal} and \eqref{sol cl}-\eqref{Temp}:
	\begin{equation}\label{T0analyt}
	T=T_0+\frac{T_0\alpha\epsilon}{C}\Big(1-J_{0}(4\omega t)\Big).
	\end{equation}
	Note, that according the asymptotic representation for the Bessel function~{\cite{Abramowitz}
	\begin{equation}\label{ass}
	\begin{split}
	&x\gg\mu+1:\\
	&J_\mu(x)=\sqrt{\dfrac{2}{\pi x}}\,\cos\left(x-\dfrac{\pi \mu }{2}-\dfrac{\pi}{4}\right)+O(x^{-3/2}),
	\end{split}
	\end{equation}
	the amplitude of the kinetic temperature oscillations decreases as $t^{-1/2}$.
	
	Fig.~\ref{ris:image1} shows
	comparison of the analytical solution~\eqref{T0analyt}
	and the numerical solution obtained by
	computer simulations of the crystal dynamics consisting of $N=5\cdot 10^4$ particles under periodic boundary conditions. In the framework of the numerical experiment, the parameters of the problem under consideration are chosen so that~$\alpha \epsilon/C=-0.1$. The simulations use the method of central differences and integration step~$0.01/\omega_e$.
	At the initial moment of time, the particle displacements are zero, and the particle velocities are random  and correspond to the crystal temperature $2T_0$. The process of energies equalization results the crystal temperature oscillates with a decreasing amplitude around $T_0$ value. The homogeneous deformation is applied when the temperature oscillations have a negligibly small amplitude. After loading of the crystal, the temperature oscillates around a new equilibrium value.
	
	In \cite{Kuzkriv2016} it is shown numerically that thermal phenomena for crystals with a sufficiently low nonlinearity do not differ greatly from harmonic crystals.
	For the simulation, a time span sufficient to describe tens of temperature oscillations after homogeneous deformation has been chosen. However, the time span is chosen not to be large so that temperature oscillations have a significant deviation from the harmonic solution.
	To calculate the mathematical expectations for the statistical quantities the results are averaged over all particles and $10^3$ realizations, which are solutions of the same equations with different random number generations for the initial conditions.
	According to work~\cite{Krivtsov 2014 DAN}, crystal temperature before the instantaneous deformation is the Bessel function of zero order. Instant deformation is applied when the amplitude of temperature oscillation is small compared to absolute value of $\alpha\epsilon/C$.

	As seen from Fig.~\ref{ris:image1} analytical solution~\eqref{T0analyt} practically coincides with the results of the numerical integration of the chain dynamics
	\mbox{equations~\eqref{deform}--\eqref{dynamic}} for several tens of oscillation periods.
	
	Formula \eqref{Tfinal} gives the limiting value for the kinetic temperature. According to expression~\eqref{Temp}, after the instantaneous deformation, the kinetic temperature oscillates around this limiting value and tends to it for the large times.
	Thus, for $t\to \infty$ expression \eqref{T0analyt} coincides with formula~\eqref{Tfinal}, obtained from the virial theorem. For arbitrary times this expression gives the desired description of the nonequilibrium transition process.

\begin{figure}
	\centering
	\includegraphics[width=0.50\textwidth, center]{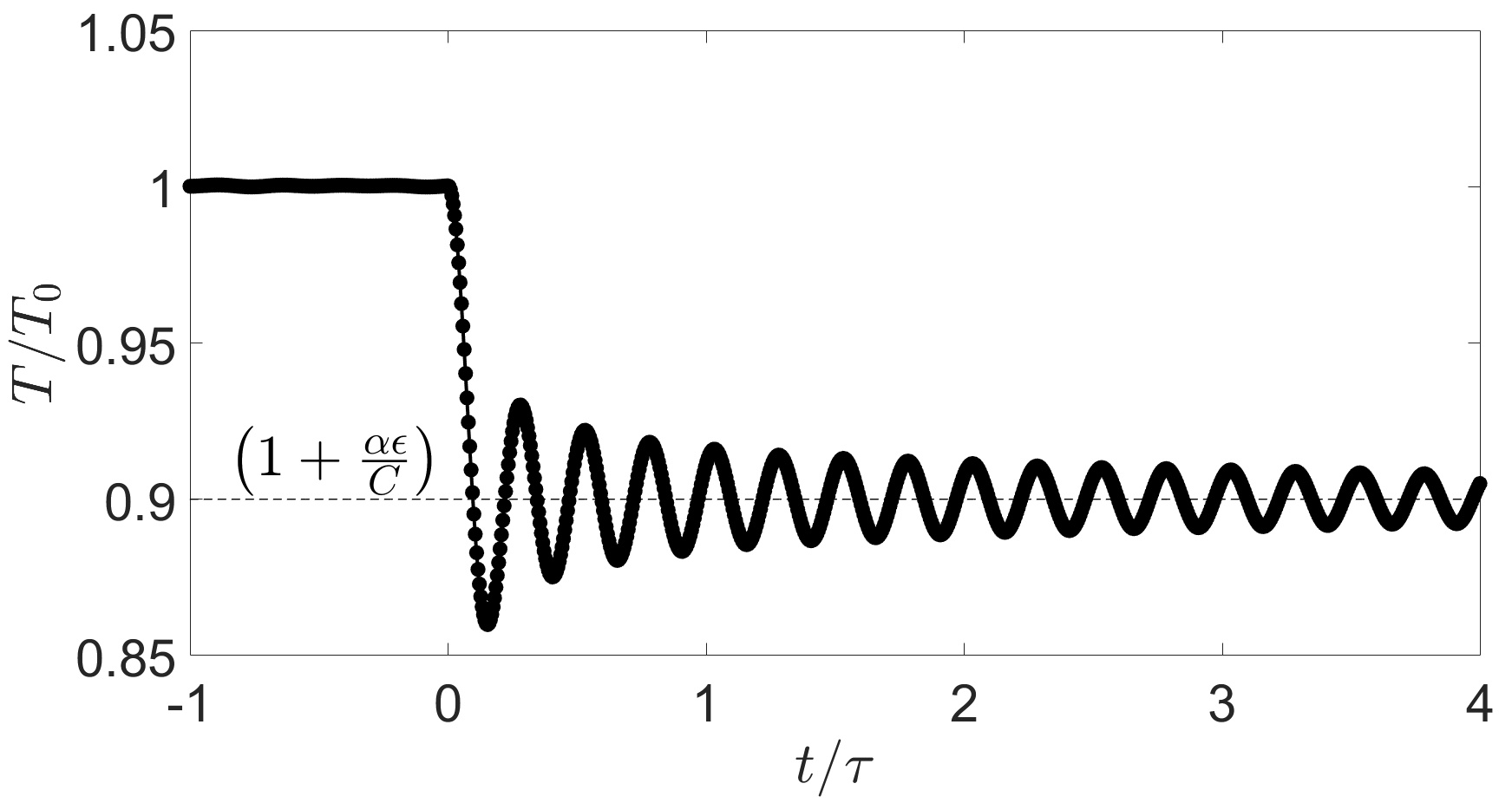}
	\caption{Oscillations of kinetic temperature $T$ in the infinite crystal after the instant loading at $t=0$. Numerical (dots) and analytical (solid line) solutions are presented. The averaging is performed using $10^3$ numerical experiments. The number of particles $N=5\cdot10^4$, constant $\tau=2\pi/\omega_e$, coefficient~$\alpha \epsilon/C=-0.1$.}
	\label{ris:image1}	
\end{figure}

	\section{Example}\label{sectEx}
		In order to estimate the temperature jump in the transition process, we consider one-dimensional ring of carbon atoms, which is in a thermodynamic equilibrium state at initial temperature~\mbox{$T_0 = 300^o$ K}. We assume that in the result of the homogeneous loading the bond deformation is $1.0$\%
		of the equilibrium interparticle \mbox{distance $a=0.154$~nm \cite{length}.}
		The mass of the carbon atom $m=1.99\cdot 10^{-26}$ kg and the first-order stiffness coefficient
		is taken equal to the stiffness of the diamond \mbox{bond
		$C=472$~N/m~\cite{Berinskii2016}.}
		 The second-order stiffness coefficient can be found from the following formula~\cite{Krivtsov2003}:
		 	\begin{equation}\label{terex}
		 \alpha=-\frac{\beta a C^2}{k_B},
		 \end{equation}
		where $\beta=0.7 \cdot 10^{-6}~{\rm K}^{-1}$
		is the coefficient of the thermal expansion of the diamond crystal~\cite{expansion}. The substitution of the
		stiffness coefficients and deformation into formula~\eqref{Tfinal} gives the new equilibrium temperature value of $298.3$ K. The asymptotic period of the kinetic temperature oscillations in the transient process described by formula \eqref{T0analyt}, is
		approximately $10.2$ femtoseconds.
		 Thus~$1.0$\% deformation leads to a change in the crystal temperature by $0.6$\%.
	
	\section{Conclusions}\label{SectConc}

{The paper presents an analytical approach for the analysis of the transition process in one-dimensional crystals (chains) subjected to an instantaneous homogeneous deformation. Such a deformation can be interpreted as an instantaneous change in the stiffness of the inter-particle bonds in a chain.
It is obtained that the transition process is accompanied by high-frequency energy oscillations, which have an analytical representation in the terms of the zero Bessel function of the first kind, consequently the amplitude of the transitional oscillations is inversely proportional to the square root of time. After the decay of the oscillations the system reaches a near equilibrium state corresponding to the predictions of equilibrium thermodynamics, however the transition process  can be studied in details with the use of the presented approach. The analytical solution is confirmed by numerical simulations. Using technique described in \cite{KuzArxiv2017} the presented approach can be extended to analyses transition processes in two-dimensional and three-dimensional materials.
The obtained results are important for establishing link between mechanical and thermal processes in solids at the femtosecond time scale.}
	
	\begin{acknowledgments}
	The authors would like to express their gratitude to professor A.\,V. Porubov, professor M.\,L. Kachanov for the useful discussions and M.\,A. Bolshanina and \mbox{N.\,D. Mushchak} for help in preparing the paper.
	\end{acknowledgments}
		
	\vspace{\baselineskip}
	This work was supported by the Russian Science Foundation (No. 19-41-04106).

\appendix
\section*{Appendixes}
\addcontentsline{toc}{section}{Appendixes}
\appendix
\section{The virial relation}\label{app0}
Following~\cite{Altenbah} the kinetic energy~\eqref{Energies} of the system can be represented by
\begin{equation}\label{Kform}
	K=\frac{m}{2}\av{u_nv_n}\dot{} - \frac{1}{2}\av{u_n( F_{n+1}-F_{n})},
\end{equation}
where expression~\eqref{dynamic} is used. The second term in~\eqref{Kform} is
\begin{equation}\label{secTerm}
	\av{u_n( F_{n+1}-F_{n})}=-\av{\varepsilon_nF_n}+\av{F_{n+1}u_n}-\av{F_{n}u_{n-1}}.
\end{equation}
Defining $g\=\av{F_{n}u_{n-1}}$, equation~\eqref{Kform}~takes the form
\begin{equation}\label{Kin}
	K=\frac{1}{2}\av{\varepsilon_nF_n}- \frac{1}{2}g'+\frac{m}{2}\av{u_nv_n}\dot{}.
\end{equation}
Values $g$ and $\av{u_nv_n}$ are constant in the thermodynamic equilibrium state and their derivatives are zero. Therefore the expression for the kinetic energy is
\begin{equation}\label{virialK}
	K=\frac{1}{2}\av{\varepsilon_nF_n}.
\end{equation}	
Formula~\eqref{virialK} allows one to obtain expressions for the equilibrium kinetic temperature of the crystal before and after the instantaneous deformation.


\begin{thebibliography}{1}

\bibitem{Goldstein2007} { Goldstein, R.\,V., Morozov, N.\,F.}
\pti{Mechanics of deformation and fracture of nano materials and nanotechnology.}
{Physical Mesomechanics}, \textbf{10}, 5-6 (2007).

\bibitem{Golovnev2006}  { Golovnev, I.\,F., Golovneva,  E.\,I.,  Fomin, V.\,M.}
\pti{ The influence of a nanocrystal size on the results of molecular-dynamics modeling.}
{Comp. Mat. Sci.} \textbf{36}, 176 (2006).

\bibitem{Korobeynikov2012}  { Korobeynikov, S.\,N., Alyokhin, V.\,V., Annin,  B.\,D., Babichev, A.\,V.}
\pti{Using Stability Analysis of Discrete Elastic Systems to Study the Buckling of Nanostructures.}
{Archives of Mechanics.} \textbf{64}, 367 (2012).

\bibitem{KrivMoroz2002} {Krivtsov, A.\,M., Morozov, N.\,F.}
\pti{ On mechanical characteristics of nanocrystals.}
Physics of the Solid State, {\bf 44}, 12 (2002).

\bibitem{Baimova2017}
{Baimova, Y.\,A., Murzaev, R.\,T., Dmitriev, S.\,V.}
\pti{ Document Mechanical properties of bulk carbon nanomaterials}
Physics of the Solid State, {\bf 56}, 10 (2014).

\bibitem{Vikesland18} Vikesland, P.\,J.
\pti{ Nanosensors for water quality monitoring}
Nature Nanotechnology, {\bf 13}, 8, (2018)

\bibitem{Hoower2015} { Hoover, W.G., Hoover, C.G.}
{ Simulation and Control of Chaotic Nonequilibrium Systems: Advanced Series in Nonlinear Dynamics: V.~27.~World Scientific, Singapore. (2015). 324 p.}

\bibitem{Altenbah} {Kuzkin, V.\,A., Krivtsov, A.\,M.} Discrete and continuum thermomechanics, Encyclopedia of Continuum Mechanics, Springer-Verlag, (2018).

\bibitem{Kuzkin2017} {Kuzkin, V.\,A.}
\pti{ Thermal equilibration in infinite harmonic crytals.}
Continuum Mech. Thermodyn., {\bf 31} (2019).

\bibitem{Bass} Bass, R.
Phys. Rev. B, {\bf 32}, 4 (1985).

\bibitem{Kuzkin2015} {Kuzkin, V.\,A., Krivtsov, A.\,M.}
\pti{Nonlinear positive/negative thermal expansion and equations of state of a chain with longitudinal and transverse vibrations.} Phys. Stat. Sol. b, {\bf 252}, 7  (2015).


\bibitem{Lepri2003}
  Lepri, S., Livi, R,  Politi, A.
  \pti{Thermal conduction in classical low-dimensional lattices.}
  Phys. Rep. {\bf 377}, 1 (2003).

\bibitem{Rieder1967}  Rieder, Z., Lebowitz, J.\,L.,  Lieb, E.,
\pti{Properties of a harmonic crystal in a stationary nonequilibrium state.}
J. Math. Phys. {\bf 8}, 5 (1967).

\bibitem{Kannan2012}
Kannan, V., Dhar, A., Lebowitz, J.\,L.
Phys. Rev. E {\bf 85}, 4 (2012).

\bibitem{Xiong2013}
Xiong, D.,  Zhang, Y.,  Zhao, H.
Phys.  Rev.  E,  {\bf 88} 5 (2013).

\bibitem{Allen1987} { Allen, M.\,P., Tildesley, A.\,K.}
\pti{ Computer Simulation of Liquids.}
Clarendon Press, Oxford. (1987). 390 p. 

\bibitem{Prig} {Klein, G., Prigogine, I.}
\pti{Sur la mecanique statistique des phenomenes irreversibles.}
Physica {\bf 19} (1953).

\bibitem{Krivtsov 2014 DAN} {Krivtsov, A.\,M.}
\pti{ Energy oscillations in a one-dimensional crystal.}
Doklady Physics, {\bf 59}, 9 (2014).

\bibitem{KuzArxiv2017} {Kuzkin, V.\,A., Krivtsov, A.\,M.}%
\pti{ Fast and slow thermal processes in harmonic scalar lattices.}
Journal of Physics: Condensed Matter, {\bf 29}, 50 (2017).

\bibitem{Kuzkriv2016} {Kuzkin, V.\,A., Krivtsov, A.\,M.}
\pti{An analytical description of transient thermal processes in harmonic crystals.}
Physics of the Solid State, {\bf 59}, 5 (2017).

\bibitem{Murachev2018}
{Murachev, A.\,S., A.M.  Krivtsov, A.\,M., Tsvetkov,  D.V.}
\pti{Thermal echo in a finite one-dimensional harmonic crystal.}
 Journal of Physics: Condensed Matter, {\bf 31}, 9 (2019).

\bibitem{GavrKriv2019}  Gavrilov, S.\,N.,  Krivtsov, A.\,M.
 Phys. Rev. E, {\bf 100}, 2 (2019).

\bibitem{Sokolov2019} Sokolov,  A.\,A., Krivtsov, A.\,M., Müller,  W.,H.,
Vilchevskaya, E.\,N. Phys. Rev. E, {\bf 99}, 4  (2019).

\bibitem{Berinskii2019} Berinskii I.\,E., Kuzkin V.\,A.
 Philosophical Transactions of the Royal Society A: Mathematical, Physical and Engineering Sciences, {\bf 378}, 2162 (2019). 

\bibitem{sen} Sederberg, S., Kong, F., Corkum, P. B. \pti{Tesla-Scale Terahertz Magnetic Impulses}. Physical Review X, 10(1) (2020).

\bibitem{Zhang} Zhang, H., Ravi-Chandar, K.
International Journal of Fracture, \textbf{142}, 3-4, (2007).

\bibitem{Morozov} Morozov, V.\,A., Petrov, Y.\,V., Sukhov, V.\,D.
Technical Physics, \textbf{64},5 (2019).


\bibitem{Kruglyak} Kruglyak, V. V., Portnoi, M. E. Generation of femtosecond current pulses using the inverse magneto-optical Faraday effect. Technical Physics Letters, {\bf 31}, 12 (2005).

\bibitem{Toyserkani} Toyserkani, E., Rasti, N., Ultrashort pulsed laser surface texturing, in Laser Surface Engineering, Edited by: J. Lawrence and D.G. Waugh, Woodhead Publishing, 718 p. (2015)

\bibitem{Tzallas} Tzallas, P., Charalambidis, D., Papadogiannis, N., Witte, K., Tsakiris, G. D. Nature, \textbf{426}, (267) (2003).

\bibitem{Orfanos}  Orfanos, I., Makos, I., Liontos, I., Skantzakis, E., Förg,  B.,  Charalambidis, D., and Tzallas, P.
\pti{Attecond pulse metrology.}
APL Photonics, \textbf{4}, (8) (2019).

\bibitem{Feng} Feng, L., Li, Y. \pti{High-intensity isolated attosecond X-ray pulse generation by using low-intensity ultraviolet–mid-infrared laser beam}. The European Physical Journal D, {\bf 72}, 9 (2018).

\bibitem{FPU_original} Fermi E., Pasta J. and Ulam S. \pti{Studies of nonlinear problems. I.} Los Alamos report LA-1940 (1955),  published later in Collected Papers of Enrico Fermi, E. Segré (Ed.), University of Chicago Press (1965).


\bibitem{FPU_review}  Berman, G. P., Izrailev, F. M. \pti{The Fermi–Pasta–Ulam problem: Fifty years of progress.} Chaos: An Interdisciplinary Journal of Nonlinear Science, {\bf 15},1 (2005).

\bibitem{Abramowitz} {Abramowitz, M. and Stegun, I.\,A.}
{ Handbook of Mathematical Functions With Formulas, Graphs, and Mathematical Tables.}
New York: Dover. (1964)  1046 p.

\bibitem{length}
Burdett Jeremy K. Chemical Bonding in Solids. New York: Oxford University Press, 1995: 152

\bibitem{Berinskii2016} {Berinskii, I.E., Krivtsov, A.M.}
\pti{ A hyperboloid structure as a mechanical model of the carbon bond.}
{International Journal of Solids and Structures. {\bf 96}, 152 (2016)}.

\bibitem{Krivtsov2003} Krivtsov A.M. {From nonlinear oscillations to equation of state in
simple discrete systems} Chaos, Solitons \& Fractals 17(1), (2003), pp. 79–87.

\bibitem {expansion}
Moelle C., Klose S., Sz$\rm \ddot u$cs F., Fecht H. J., Johnston C., Chalker P. R., Werner M.
\pti{Measurement and calculation of the thermal expansion coefficient of diamond.}
Diamond and Related Materials, {\bf 6}(5-7) (1997)

			
		\end{thebibliography}
	\end{document}